\documentclass[aps,twocolumn]{revtex4}
\usepackage{epsfig}

\begin{document}

\title{On the trapping reaction with mobile traps}

\author{Vishal Mehra and Peter Grassberger}

\affiliation{John-von-Neumann Institute for Computing, Forschungszentrum J\"ulich,
D-52425 J\"ulich, Germany}

\date{\today}

\begin{abstract}
We present Monte Carlo results for the two-species trapping reaction $A+B \to B$
with diffusing $A$ and $B$ on lattices in one, two and three dimension. We use 
a novel algorithm which permits to simulate survival probabilities of $A$ particles
down to $<10^{-30}$ with high accuracy. The results for the survival probability
agree much better with the exact asymptotic predictions of Bramson and Lebowitz 
(Phys.  Rev. Lett.  {\bf 61}, 2397 (1988)) than with the heuristics of 
Kang and Redner (J.  Phys. A {\bf 17}, L451 (1984)). But there are very large 
deviations from either which show that even these simulations are far from 
asymptotia. This is supported by the rms. displacement of $A$ particles which 
clearly show that the asymptotic regime has not been reached, at least for 
$d=2$ and $d=3$.
\end{abstract}

\maketitle

Even the simplest reaction diffusion systems with only two-body reactions and 
without particle production show very rich and not yet fully understood phenomena. 
One prototype is the absorption of diffusing particles on randomly located
static sinks. This model, studied first by Smoluchowski nearly hundred years ago
\cite{smoluch}, shows an exponential decrease of the particle density in mean 
field theory, but the exact asymptotic behaviour found by Donsker and Varadhan 
\cite{dv75} is a stretched 
exponential in any finite dimension $d$. Another classic example, the 
recombination $A + A \to A$, was studied by Feller \cite{feller}. It leads to 
$n_A \sim t^{-1/2}$ in $d=1$, as compared to the mean field solution 
$n_A \sim t^{-1}$ which holds for $d>2$.

Particle annihilation with two mobile species $A$ and $B$, either of the type 
$A+B\to B$ or $A+B\to 0$, were studied by Ovchinikov and Zeldovich \cite{oz78}
and Toussaint and Wilczek \cite{tw83}. The motivation of \cite{tw83} stemmed 
from the fate of magnetic monopoles in the early universe, but applications to 
condensed matter physics and chemical kinetics are of course more numerous.

If one starts out with equal densities for $A$ and $B$, $n_A(0) = n_B(0)$, then 
the reaction $A+B\to 0$ leads to $n \sim t^{-1/4}$ \cite{tw83}. This is 
different from the cases $n_A(0) < n_B(0)$ and $A+B\to B$. In both these cases 
$n_A \ll n_B$ for late times, and one expects identical asymptotics. This
was indeed proven rigorously by Bramson and Lebowitz \cite{bl88} who obtained
\begin{equation}
   n_A(t) \sim \left\{ \begin{array}{ll}
               \exp(-\lambda_1\sqrt{t})  \qquad & d=1 \\
               \exp(-\lambda_2 t/\ln t)  \qquad & d=2 \\
               \exp(-\lambda_d t)        \qquad & d\ge3
               \end{array}   \right.
                          \label{br-le}
\end{equation}
with unknown constants $\lambda_d$. In contrast to this, different asymptotics
had been predicted for $A+B\to B$ and for $A+B\to 0$ with $n_A(0) < n_B(0)$ by
Kang and Redner \cite{kr84,rk84} by heuristic arguments. 

Verifying Eq.(\ref{br-le}) numerically has turned out to be about as difficult
as verifying the Donsker-Varadhan stretched exponential. The first simulations 
by Kang and Redner \cite{kr84} agreed with their own (supposedly wrong)
heuristic asymptotics.
Subsequent simulations \cite{ssb90,ssb91} were judged even by the authors as 
inconclusive. The main problem seems to be that there are large finite time
corrections to the asymptotic behaviour. Thus one would like to simulate up 
to very long times. But with the straightforward approaches used so far it 
is practically impossible to estimate survival probabilities smaller than 
$\approx 10^{-8}$, even with the most powerful present day computers.

It is the purpose of this work to present simulations for the reaction 
$A+B\to B$ which go far beyond this. We shall not give any results for 
$A+B\to 0$ since our special numerical methods cannot be applied to that case.
More specifically we consider regular lattices on which the $B$ particles
perform independent random walks. $A$ particles also perform random walks until
they hit upon a $B$ particle in which case they are instantaneously absorbed.
Initial conditions are such that all $B$s are uncorrelated with homogeneous 
concentration $c$. Since $A$ particles do not interact with each other, their
concentration is irrelevant. In the actual simulations we shall use either
one or two $A$s in the initial configuration. Notice that in this model the 
$B$ particles act as catalyzers, and their distribution is Poissonian at 
any time, if they start out independently at $t=0$.

Our algorithm is related to an algorithm used recently \cite{mg01} for the 
trapping (Donsker-Varadhan) problem. There we were able to clarify the cross-over
from the mean field type to the stretched exponential behaviour. In the 
present case we shall see that -- despite going to much longer times than in 
earlier simulations -- we still do not yet fully understand the cross-over 
to the Bramson-Lebowitz asymptotics. 

The algorithm has several essential ingredients. The first is that we use cloning
(``enrichment") of configurations with surviving $A$ particles and a bias 
for the diffusion of $A$ such that less of them are absorbed. This bias is 
compensated by suitably chosen weights, i.e. we always deal with nontrivially 
weighted ensembles. On the 
other hand, the simulation is stopped as soon as one of the $A$s is absorbed,
or if the weight of the configuration is too
small. Thus all our results are based on {\it conditional probability 
distributions}, conditioned on the survival of all $A$s. These features 
are implemented by PERM \cite{g97} which is a general growth method (using
`sequential importance sampling with resampling' in the sense of \cite{liu01})
and has been very successful in a large number of problems \cite{mg01,gn01}.

Assume that we have a single $A$ (the case for $k$ $A$ particles with $k>1$
is straightforward) which has arrived at site $i$ at time $t$. The fact that we 
condition on those events where $A$ is not absorbed means that there 
cannot be a $B$ at this site. Thus the homogeneity of the unconditioned 
distribution is broken, and an effective {\it hole} in the $B$ distribution
is introduced. For times $>t$ this hole makes a random walk. If it meets a 
hole produced at a time $\neq t$, the two recombine.

In principle one could simulate these holes explicitly, i.e. one could 
simulate the $A$s in a background of $B$ where a $B$ particle (sic!) is removed 
each time it hits an $A$. We indeed did perform such simulations. They agreed
with straightforward simulations not using PERM or any conditioning and were
more accurate, but the accuracy of both type of simulations was 
rather poor in comparison with our final algorithm. 

Our final ingredient is that we do an exact summation over the 
$B$ paths. This depends crucially on the fact that the $B$ distribution 
is Poissonian. The latter is still true for the conditioned distribution
(it would not be true in the reaction $A+B\to 0$, therefore our method cannot 
be used for it). A Poisson distribution is uniquely characterized by its 
mean. The evolution of this mean is described by a modified diffusion equation.
More precisely, we describe the $B$ density by a homogeneous background $c$ 
minus a density $\rho(i,t)$ of holes. The latter is at every time step 
set equal to $c$ at the actual $A$ position, but otherwise 
evolves according to the simple diffusion equation $\rho(i,t+1) = (2d)^{-1}
\sum_{<j,i>} \rho(j,t)$. Its initial condition is $\rho(i,0)=0$.
At each time step the $A$ (which is assumed to be at site $i$) has a chance 
$\exp(\rho(i,t)-c)$ to be absorbed, i.e. the weight of the event in the 
conditioned ensemble decreases by a factor $\exp(c-\rho(i,t))$. 
The algorithm used in \cite{mg01} for the trapping problem is just a 
simplification where we omit the diffusion of the holes.

All our simulations were done on workstations, with a total of a few hundred 
hours of CPU time. All results were carefully checked for small $t$ against 
straightforward brute force simulations with the most simple algorithm. In
addition we also made simulations with algorithms of intermediate complexity
and efficiency. 

We first discuss results for $d=1$. Here we can use lattices so 
large that none of the holes ever reaches the boundary, thus we have 
no finite size effects at all. In Fig.1a we show the survival chances 
$P_A(t)$ of a single $A$ particle and of a pair of particles which started 
at the same site. Although they become as small as $10^{-35}$,
the statistical errors are much smaller than the thickness of the lines. 
The fact that $P_A$ for a pair is larger that the square of $P_A$ for a 
single particle is easily understood: If already one $A$ particle has survived in some
region, there are less than average $B$ particles in this region, and the 
second $A$ particle has a bigger survival chance.
Theoretically \cite{bl88} we expect $P_A(t)\sim\exp(-const\sqrt{t})$, at 
least for a single particle. In Fig.1b we thus show $\sqrt{t} \ln P_A(t)$ 
on a semi-logarithmic 
scale. Error bars are here $< 0.001$. Thus the decline for $t> 1000$ is 
statistically highly significant for both curves. Unless we accept that 
there is an error in 
\cite{bl88}, we have to conclude that at least this decline of $-\sqrt{t} \ln P_A(t)$ 
for a single $A$ particle 
is a finite $t$ effect, and that the true asymptotic behaviour sets in at
$t \gg 10^4$. We also compared our data with the prediction $-\ln P_A(t) 
\sim t^{-1/4}$ of \cite{kr84}, with even worse agreement. Although that 
latter prediction was for $A+B\to B,\; n_A(0) < n_B(0)$, we pointed out 
already that both models should show the same asymptotics.
We thus conclude that the data are in rough agreement with \cite{bl88},
but the very big deviations are surprising (in particular since they are 
not monotonic!) and not understood.

\begin{figure}
%Fig 1a,b
\psfig{file=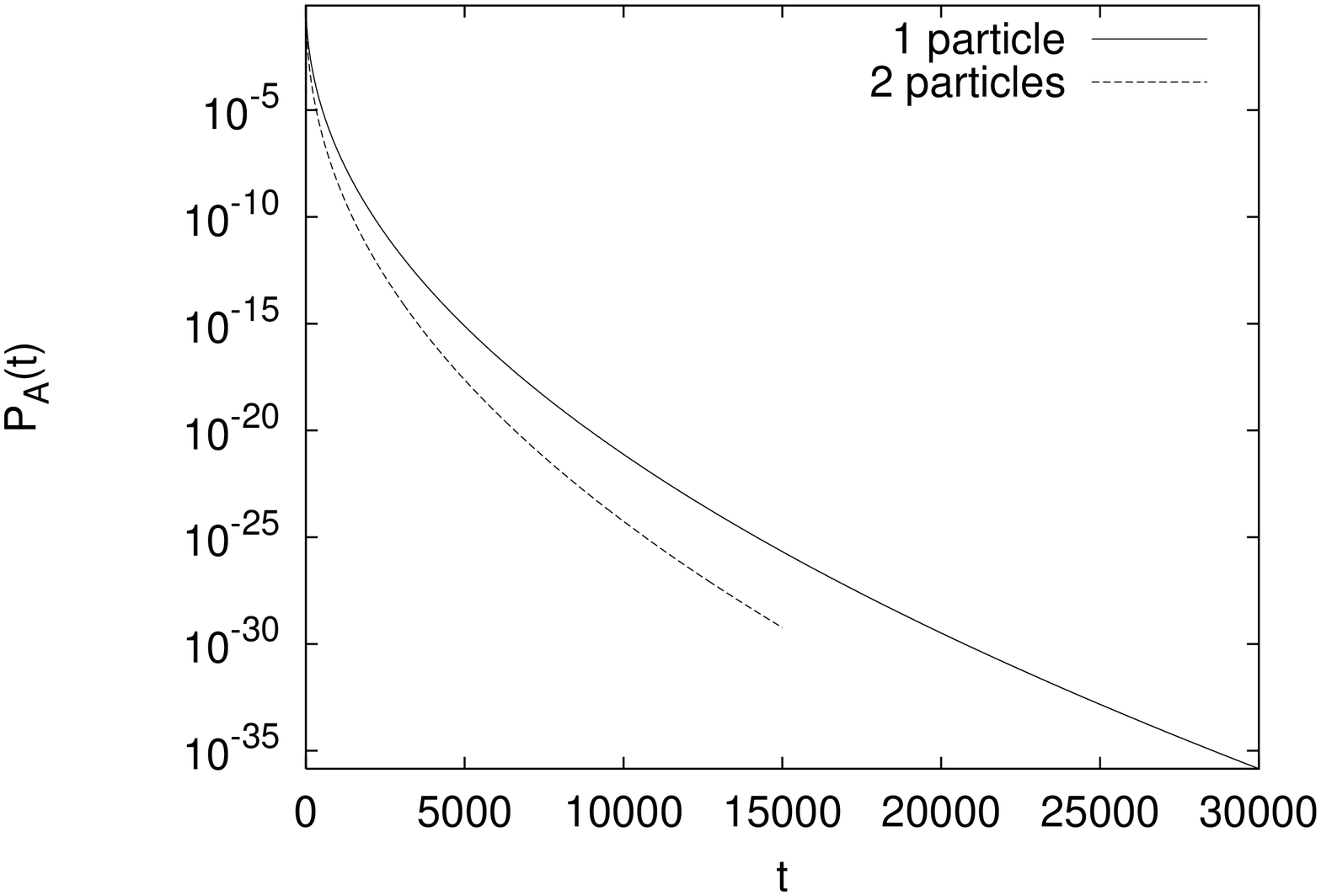,width=5.8cm,angle=270}
\psfig{file=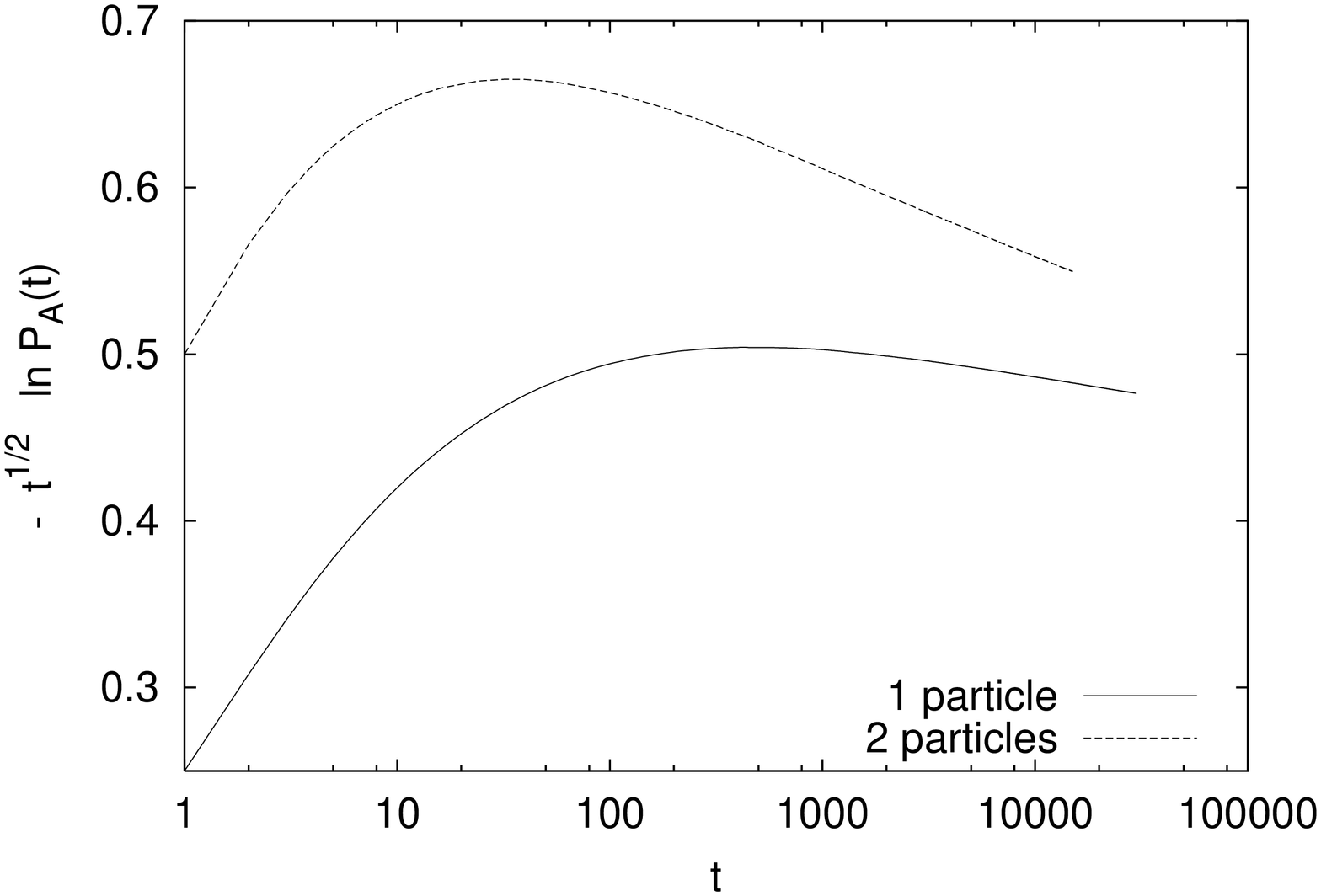,width=6.cm,angle=270}
\caption{(a) Semi-logarithmic plot of the survival probability in $d=1$ for 
   $c=0.5$. The upper curve is for a single $A$ particle, the lower for a 
   pair. \\ \noindent
   (b) $-\sqrt{t} \ln P_A(t)$ from the same data as in panel (a), plotted 
   on a logarithmic horizontal scale.}
\label{1D_1p_pob}
\end{figure}

The mean squared displacement $\langle R^{2}_A\rangle$ of the $A$ particle 
is shown in Fig.2. In this figure we also show the squared distance between 
two $A$ particles which started off simultaneously at the same site. We see that 
$\Delta R$ increases less fast than $R$, indicating an effective attraction 
mediated by the $B$ particles: If already one $A$ particle has survivied in some 
region, there are less than average $B$ particles in this region, and the 
second $A$ particle will not only survive longer, but it will also 
survive preferentially in a region close to the first one. But this 
appears to be a weak effect. The most economic conclusion from Fig.2 is
that both curves converge to the same scaling behaviour,
$R^2 \sim (\Delta R)^2 \sim t^\nu$ with $\nu = 0.5$ to $0.6$.  But any 
determination of a critical index should be taken with very big caution in 
view of Fig.1: we should not take any behaviour seen for $t<10^5$ as reflecting
the asymptotic behaviour.

\begin{figure}
%Fig 2
\psfig{file=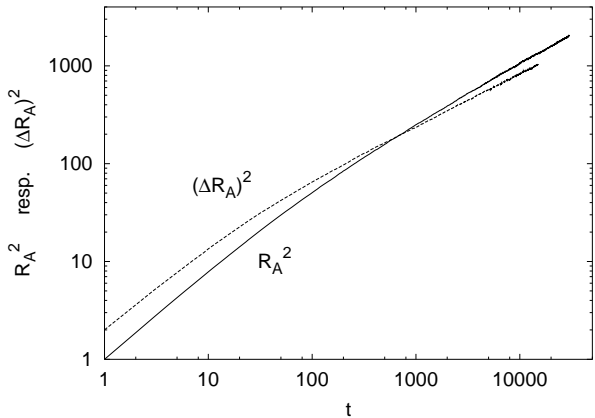,width=5.8cm,angle=270}
\caption{Full line: mean squared displacement of the $A$ particle in $d=1$, 
   for $c=0.5$. Broken line: mean squared distance between two $A$ particles,
   when conditioned on survival of both.}
\label{1D_1R}
\end{figure}

\begin{figure}
%Fig 3
\psfig{file=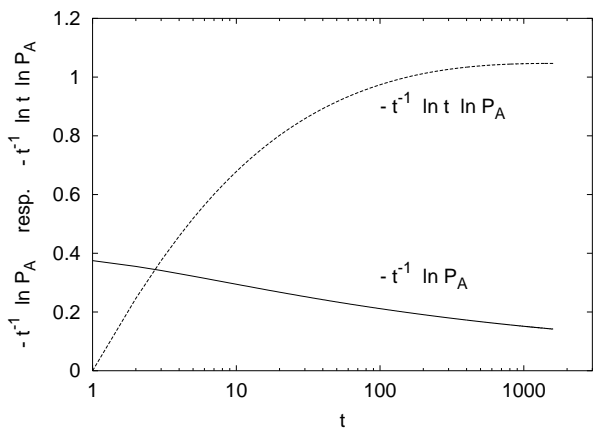,width=5.8cm,angle=270}
\caption{Full line: $-t^{-1} \ln P_A(t)$ for $d=2$ and $c=0.5$; Broken line:
   $-t^{-1} \ln t \ln P_A(t)$ from the same data. The horizontal scale is 
   logarithmic, the vertical is linear.}
\label{2d-P}
\end{figure}

With increasing $d$, updating $\rho(i,t)$ becomes more and more time consuming. 
Thus we can simulate only for much shorter times, if we want to avoid excessive
CPU times or large finite size corrections. For $d=2$, results for 
$t^{-1} \ln P_A(t)$ with $c=0.5$ are shown in Fig.3, together with 
$
   - t^{-1}\; \ln t \;\ln P_A(t)\;.
$
For large values of $t$ the data agree much better with Eq.(\ref{br-le}) than with 
the alternative prediction $-\ln P_A(t) \sim at - b t^{1/2}$ of \cite{rk84}.
The factor $1/\ln t$ in Eq.(\ref{br-le}) seems to be correct asymptotically, although
it makes agreement worse for small $t$. Anyhow, deviations from 
Eq.(\ref{br-le}) are substantial, and even for the largest $t$ where the curve
appears to be horizontal in the figure (and where, by 
the way, $P_A(t)\approx 10^{-98}$) it still shows a definite curvature.

The fact that asymptotia cannot have been reached by these 2-d data is 
most clearly seen from $\langle R^{2}_A\rangle$. It is shown in Fig.4, 
together with the analogous data from $d=1$ and $d=3$. To make the point
particularly clear we show there $\langle R^{2}_A\rangle / \sqrt{t}$ against 
$t$. If $d=2$ is the upper critical dimension for $A+B\to B$ as suggested 
by Eq.(\ref{br-le}), then we should expect $\langle R^{2}_A\rangle \sim t$ 
up to logarithmic corrections in $d=2$. We also should expect that the data 
for $d=2$ should fall between those for $d=3$ (where we expect 
$\langle R^{2}_A\rangle \sim t$) and $d=1$ (where $\langle R^{2}_A\rangle 
\sim t^\nu$ with $\nu < 1$). Figure 4 shows a completely different
behaviour. On the one hand, $\langle R^{2}_A\rangle$ is {\it very far} from 
being $\sim t$, on the other hand the data are not monotonic with $d$.

\begin{figure}
%Fig 4
\label{1p_rsq}
\psfig{file=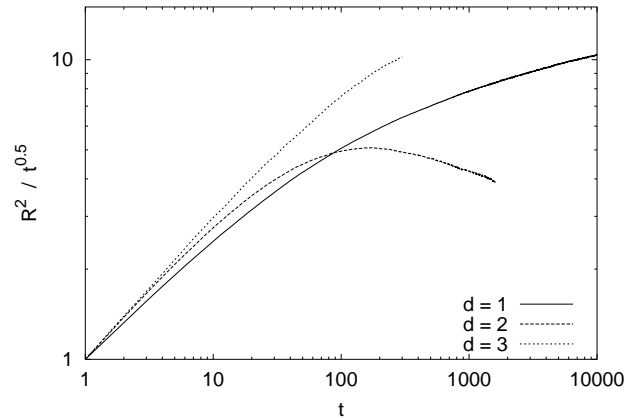,width=5.8cm,angle=270}
\caption{Log-log plot of mean squared displacements in 1, 2 and 3 dimensions,
    divided by $\sqrt{t}$. }
\end{figure}

One might expect that asymptotic behaviour is observed earlier for higher
values of the concentration $c$. We performed therefore also 2-d simulations 
with $c=0.7$ and $c=1.0$. As expected, $P_A$ and $\langle R^{2}_A\rangle$
both decreased with $c$, but the strange time dependence of 
$\langle R^{2}_A\rangle$ persisted.

\begin{figure}
%Fig 5
\label{3p_fig}
\psfig{file=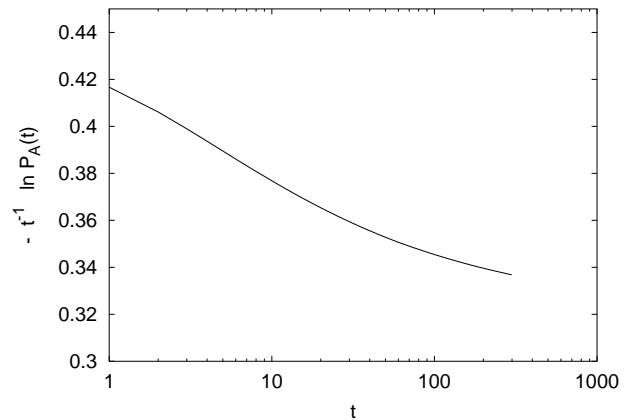,width=5.8cm,angle=270}
\caption{Log-linear plot of $-t^{-1} P_A(t)$ for $d=3,\; c=0.5$.}
\end{figure}

Finally, we show in Fig.5 the survival probability in $d=3$, again for 
$c=0.5$. More precisely, we show $-\ln P_A(t)/ t$. It depends quite 
strongly on $t$, but its curvature is consistent with convergence to 
Eq.(\ref{br-le}) without further surprises.

In conclusion, we have performed simulations of a trapping model in which both 
traps and trapped particles are mobile with equal mobilities. The simulations, 
undertaken in the hope of understanding the precise cross-over to the exactly
known asymptotic behaviour, are much more extensive than previous simulations
of this model, in the sense of reaching much longer times and much lower 
survival probabilities. This was possible due to a novel algorithm. 
In spite of the vastly improved numerics we have not been able to understand 
all details of the model. Some of our results are indeed quite puzzling.
On the other hand, our methods can be possibly applied also to other 
models where absorbers are free particles undisturbed by the particles which 
they absorb. One class of such models are e.g. gated reaction-diffusion 
systems of the type $A+B\to 0$ with stochastically changing reaction rates
\cite{bmo00}.

Acknowledgement: We thank Walter Nadler for discussions and a careful reading of the 
manuscript.


\begin{thebibliography}{99}
\vglue -.5cm
\bibitem{smoluch} M.V. Smoluchowski, Phys. Z. {\bf 17}, 557, 585 (1916).
\bibitem{dv75} M. Donsker and S.R.S. Varadhan, Commun. Pure Appl. Math. {\bf
   28}, 525 (1975); {\bf 32}, 721 (1979).
\bibitem{feller} W. Feller, {\it An Introduction to Probability Theory and 
   its Applications}, v. 1, 3rd ed., Wiley, New York (1968).
\bibitem{oz78} A.A. Ovchinikov and Ya.B. Zeldovich, Chem. Phys. {\bf 28}, 214 
   (1978).
\bibitem{tw83} D. Toussaint and F. Wilczek, J.  Chem.  Phys.  {\bf 78}, 
   2642 (1983).
\bibitem{bl88} M. Bramson and J.L. Lebowitz, Phys.  Rev.  Lett.  {\bf 
   61}, 2397 (1988); J. Stat. Phys. {\bf 62}, 297 (1991). 
\bibitem{kr84} K. Kang and S. Redner, Phys. Rev. Lett. {\bf 52}, 955 (1984).
\bibitem{rk84} S. Redner and K. Kang, J.  Phys. A {\bf 17}, L451 (1984).
\bibitem{ssb90} H. Schn\"orer, I.M.  Sokolov, and A. Blumen, Phys.  Rev.
   A {\bf 42}, 7075 (1990).
\bibitem{ssb91} I.M.  Sokolov, H. Schn\"orer, and A. Blumen, Phys.  Rev. 
   A {\bf 44}, 2388 (1991). 
\bibitem{mg01} V. Mehra and P. Grassberger, cond-mat/0107525 (2001).
\bibitem{g97} P. Grassberger, Phys. Rev. E {\bf 56}, 3682 (1997).
\bibitem{liu01} J.S. Liu, {\it Monte Carlo Strategies in Scientific Computing},
   Springer Series in Statistics (Springer, New York 2001)
\bibitem{gn01} P. Grassberger and W. Nadler, cond-mat/0010265 (2000).
\bibitem{bmo00} O. Benichou, M.  Moreau and G.  Oshanin, Phys.  Rev.  E 
{\bf 61}, 3388 (2000). 

\end{thebibliography}
\end{document}